\documentclass[twocolumn,showpacs,preprintnumbers,amsmath,amssymb,superscriptaddress]{revtex4-1}
\usepackage{graphicx}
\usepackage{dcolumn}
\usepackage{bm}
\usepackage{color}
\usepackage{float}
\usepackage[english]{babel}
\usepackage[utf8x]{inputenc}
\usepackage[T1]{fontenc}

\usepackage{amsmath}
\usepackage[colorinlistoftodos]{todonotes}
\usepackage[colorlinks=true, allcolors=blue]{hyperref}
\usepackage{textgreek}

\begin{document}

\title{A simple approach to bulk bioinspired tough ceramics}

\author{Hassan Saad}
\affiliation{Laboratoire de Synthèse \& Fonctionnalisation des Céramiques, UMR3080 CNRS-Saint-Gobain CREE, Saint-Gobain Research Provence, 84306 Cavaillon, France}
\affiliation{Univ Lyon, INSA Lyon, MATEIS CNRS UMR5510, Villeurbanne, France}
\author{Kaoutar Radi}
\affiliation{Univ. Grenoble Alpes, CNRS, Grenoble INP, SIMaP, Grenoble, France}
\author{Thierry Douillard}
\affiliation{Univ Lyon, INSA Lyon, MATEIS CNRS UMR5510, Villeurbanne, France}
\author{David Jauffres}
\affiliation{Univ. Grenoble Alpes, CNRS, Grenoble INP, SIMaP, Grenoble, France}
\author{Christophe L. Martin}
\affiliation{Univ. Grenoble Alpes, CNRS, Grenoble INP, SIMaP, Grenoble, France}
\author{Sylvain Meille}
\affiliation{Univ Lyon, INSA Lyon, MATEIS CNRS UMR5510, Villeurbanne, France}
\author{Sylvain Deville}
\email{Corresponding author: sylvain.deville@univ-lyon1.fr}
\affiliation{Laboratoire de Synthèse \& Fonctionnalisation des Céramiques, UMR3080 CNRS-Saint-Gobain CREE, Saint-Gobain Research Provence, 84306 Cavaillon, France}
\affiliation{now with: Université de Lyon, Université Claude Bernard Lyon 1, CNRS, Institut Lumière Matière, 69622 Villeurbanne, France}

\date{\today}

\begin{abstract}
The development of damage-resistant structural materials that can withstand harsh environments is a major issue in materials science and engineering. Bioinspired brick-and-mortar designs have recently demonstrated a range of interesting mechanical properties in proof-of-concept studies. However, reproducibility and scalability issues associated with the actual processing routes have impeded further developments and industrialization of such materials. Here we demonstrate a simple approach based on uniaxial pressing and field assisted sintering of commercially available raw materials to process bioinspired ceramic/ceramic composites of larger thickness than previous approaches, with a sample thickness up to 1~cm. The ceramic composite retains the strength typical of dense alumina (430~$\pm$~30~MPa) while keeping the excellent damage resistance demonstrated previously at the millimeter scale with a crack initiation toughness of 6.6~MPa.m$^{1/2}$ and fracture toughness up to 17.6~MPa.m$^{1/2}$. These results validate the potential of these all-ceramic composites, previously demonstrated at lab scale only, and could enable their optimization, scale-up, and industrialization. 

\end{abstract}

\maketitle

The strong and directional iono-covalent bondings confer to ceramics their high strength, high stiffness and their important thermal stability which make them good candidates for structural and high temperature applications~\cite{Ashby2011}. However, this atomistic configuration explains their lack of dislocation-induced plasticity and in turn their low tolerance to defects and their brittleness, which limits their use for applications where resistance to fracture is required~\cite{Ritchie2011}. To improve the damage resistance of engineering materials, lessons have been taken from structural materials found in nature. Nacre, a school-case for the design of tough ceramic composites~\cite{Wegst2015}, is defined by several microstructural features at different scales, inducing several toughening mechanisms such as crack deflection, bridging and platelet pull-out~\cite{Barthelat2010, Barthelat2011}. Implementing such designs in engineering materials actually used in applications poses scientific, technical, and industrial challenges unaddressed so far.

Materials inspired by this design are referred as brick-and-mortar materials, and many different approaches have been reported so far, each coming with its own set of advantages and limitations. Building upon lessons learned from more traditional composite materials, the introduction of a secondary soft phase, usually a polymer or a metal, results in composites with varying degrees of damage resistance. Such materials, however, because they comprise an organic or metallic phase, cannot be used for high temperature applications or in aggressive chemical environments~\cite{Deville2006, Munch2008, Launey2009a, Ferraro2019}. An alternative solution to obtain brick-and-mortar microstructures is to align platelets (bricks) using an external driving force while controlling the interphase (mortar) composition. The commercial availability of monocrystalline alumina platelets recently expanded the range of possibilities, and lead to a new type of ceramic/ceramic composites~\cite{Bouville2014}. Several strategies have been reported to align such platelets, including ice-templating~\cite{Bouville2014}, the combined use of magnetic field and slip casting~\cite{LeFerrand2015,LeFerrand2019,Pelissari2018, Pelissari2019}, or additive manufacturing (robocasting)~\cite{Feilden2017a}.

In addition to reproducibility issues, a common limitation of all these techniques is the size of samples obtained and the potential for scale-up and industrialization, which has not been addressed yet. The thickness of samples produced so far in academic labs does not exceed 2--3~mm. The development of high performance structural materials relies today not only on reaching the desired mechanical properties, but also on the ability to develop scalable, industrializable, and cost-effective processing routes. It was shown recently~\cite{Evers2019} that the uniaxial pressure applied during field-assisted sintering (FAST) was sufficient to induce platelet alignment at the microscale, yielding high energy absorption abilities. Although densities up to 99~\% can be obtained, the final microstructures always exhibited important grain growth.

Here we demonstrate that FAST combined with uniaxial pressure is suitable to obtain centimeter-thick samples with a nacre-like microstructure, starting from an alumina platelets system containing alumina nanoparticles and glass-phase precursors. We report for the first time strength and toughness ($R-$curves) on cm-thick nacre-like ceramic composites, validating the mechanical properties demonstrated previously on millimeter-size samples only~\cite{Bouville2014}. %

\begin{figure*}[ht]
\centering
\includegraphics[width=17.5cm]{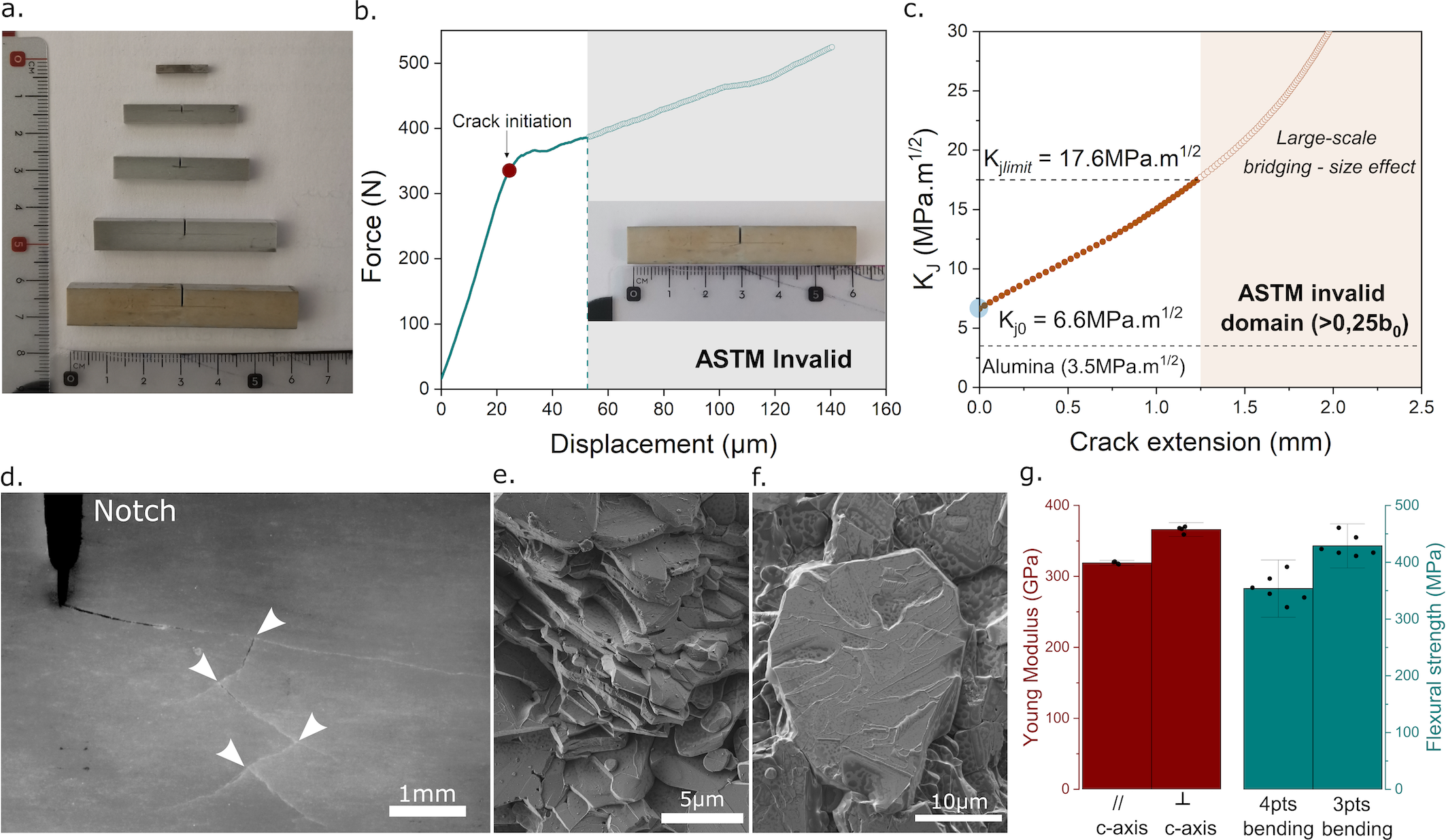}
\caption{\textbf{Mechanical characterization and fracture surfaces.} (a) Centimeter-sized specimen with 2~mm, 5~mm, 6~mm, 8~mm, and 10~mm thickness (b) Force-displacement curves obtained from SENB tests for the $5\times10\times60$~mm$^3$ geometry. Full circle indicates the crack initiation ; (c) $R-$curve : toughness versus crack extension measured on the $5\times10\times60$~mm$^3$ sample (1~cm thickness). Hollow markers correspond to the points of the curve beyond the ASTM-E1820 standard limit; (d) Surface of SENB sample showing crack initiation from the notch plane and a long-range crack propagation with multiple crack deviations (white arrows); (e) Fracture surface showing a stack of well-aligned platelets (nacre-like microstructure); (f) Presence of the aluminosilicate glass-phase at the platelets surface; (g) Mechanical properties of the material : Young's modulus as a function of the platelets orientation and flexural strength ($3\times6\times36$~mm$^3$ samples).  \textsuperscript{\textcopyright} (2020) H. Saad \emph{et al.} (10.6084/m9.figshare.11907276) CC BY 4.0 license https://creativecommons.org/licenses/by/4.0/.}
\label{fig:SENB_bulk}
\end{figure*}

\textbf{A simple and scalable processing route} -- The system used in this work was described by \textit{Bouville et al.} and is composed of anisotropic alumina platelets, alumina nanoparticles and glass-phase precursors ($CaCO_3$, $SiO_2$)~\cite{Bouville2014}, used respectively for the formation of the mineral bridges and the glass-phase between the platelets. We considerably simplify the process compared to previous studies. An aqueous suspension with all the particles and precursors is homogeneously dispersed, quenched in liquid nitrogen to retain the spatial repartition of the different elements, and freeze-dried. The dried powder is thermally treated to eliminate the remaining organic compounds (dispersant and binder) and initiate the decomposition of calcium carbonate to calcium oxide. The powder is then introduced into a die to prepare samples with thickness up to 1~cm (after sintering). No pre-alignment step is performed prior to sintering which enables the preparation of thick samples (Figure~\ref{fig:SENB_bulk}a). The material relative density after sintering was evaluated at 98~\% (3.88~g.cm$^{-3}$). The uniaxial pressure applied during FAST align the platelets while the high heating rate limits grain growth, ensuring that a dense and fine microstructure is retained.

\textbf{Strong and tough alumina of high thickness} -- Single-edge notched beam (SENB) tests were used to evaluate the material toughness during stable crack propagation performed on 1~cm thick samples and following the ASTM-E1820 standard~\cite{ASTME1820}. Crack initiation can be directly observed from the \textit{force-displacement} curve (Figure~\ref{fig:SENB_bulk}b) when it starts to deviate from the initial linear elastic part. This is followed by an important reinforcement domain characterized by a long and stable crack propagation.

The damage resistance of the material is evaluated by measuring the $R-$curve, the evolution of the apparent toughness $K_J$ during stable crack propagation (Figure~\ref{fig:SENB_bulk}c). A crack initiates consistently at 80~$^{\circ}$ from the notch plane and propagates in a stable manner during the test (Figure~\ref{fig:SENB_bulk}d). The beginning of the curve corresponds to the apparent $K_J$ fracture toughness of the material at crack initiation $K_{J0}$, evaluated here to 6.6~MPa.m$^{1/2}$. This value is consistent to that obtained using the same composition but processed by ice-templating and FAST (6.1~MPa.m$^{1/2}$)~\cite{Bouville2014} and is the double of that of a standard polycrystalline alumina (3.5~MPa.m$^{1/2}$). The maximum valid toughness according to ASTM-E1820 standard is 17.6~MPa.m$^{1/2}$.

For the first time a $R-$curve has been obtained on a centimeter-sized dense brick-and-mortar ceramic composite when all the studies published so far were made with samples of thickness not exceeding 2--3~mm~\cite{Bouville2014,LeFerrand2015,Pelissari2018,Munch2008}. The interest of working with thicker samples, besides the obvious potential for industrial applications, is to delay the apparition of size-effects and to increase the characterization domain of the toughening mechanisms. Post-mortem observations of the sample fracture surface confirm the presence of toughening mechanisms such as crack deflection and crack bridging (Figure~\ref{fig:SENB_bulk}d), while figure~\ref{fig:SENB_bulk}e reveals the local arrangement of platelets and the nacre-like microstructure. Some breakage of platelets that occurred during crack propagation can be observed, as well as the presence of the aluminosilicate glass-phase at the surface of platelets (Figure~\ref{fig:SENB_bulk}f). 

The Young's modulus $E$ and Poisson's ratio $\nu$ of the material (Fig.~\ref{fig:SENB_bulk}g) were found to be respectively $E$=368~GPa and $\nu$=0.24 along the platelets thickness (parallel to $c$-axis of the alumina platelets), and $E$=320~GPa along the platelets length (perpendicular to $c$-axis), the value of $\nu$ remaining the same, which is consistent with the standard values for polycrystalline alumina (300-400~GPa). 

The mechanical resistance was evaluated using 4-points bending (4PB) tests, perpendicularly to the platelets preferential direction ($\perp$ to their thickness). The material exhibited a linear elastic behavior up to fracture with an average flexural strength of 350~$\pm$~40~MPa, typical of dense polycrystalline alumina (300--400~MPa). 3-points bending (3PB) tests were also performed on specimens of similar size yielding a flexural strength of 430~$\pm$~30~MPa (Figure~\ref{fig:SENB_bulk}g). The difference may be explained by the larger volume under tensile stress in 4PB than in 3PB, leading to a lower strength in 4PB. 

\begin{figure*}
\centering
\includegraphics[width=12cm]{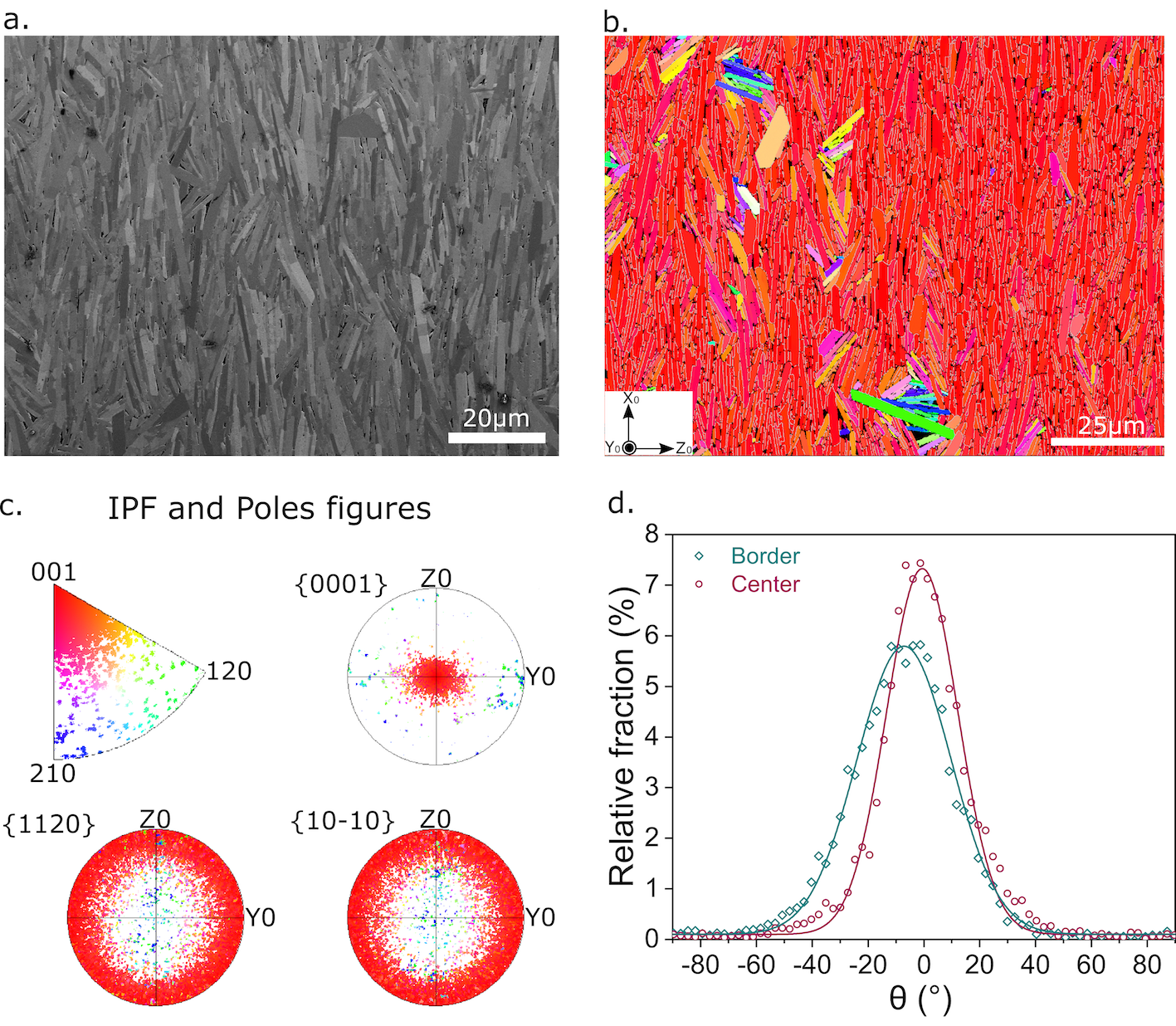}
\caption{\textbf{Microstructure of the nacre-like alumina}. (a) Cross-section micrograph of the sample at low magnification showing a preferential alignment of the platelets perpendicularly to the applied pressure (scale bar: 20{\textmu}m). The nacre-like microstructure is fine and dense. The uniaxial pressing direction (Z$_0$) is horizontal (b) Quantification of platelets orientation. ESBD map of a cross section (center of the sample). c) EBSD pole figure and inverse pole figure (IPF). The platelets are oriented such that their $c$-axis is parallel to the pressing direction Z$_0$. d) Platelets distribution as a function of the angle $\theta$. Grain misalignment is evaluated using full-width at half maximum (FWHM) of the distribution curve. \textsuperscript{\textcopyright} (2020) H. Saad \emph{et al.} (10.6084/m9.figshare.11907276) CC BY 4.0 license https://creativecommons.org/licenses/by/4.0/.}
\label{fig:EBSD}
\end{figure*}

\textbf{A nacre-like microstructure} -- A dense and highly textured nacre-like microstructure (Figure~\ref{fig:EBSD}a) explains these excellent mechanical properties, with a platelet preferential direction perpendicular to the applied pressure. The platelet orientation was evaluated using electron backscattered diffraction (EBSD). The map in figure~\ref{fig:EBSD}b was performed at the center of the sample and confirms the texture. According to the inverse pole figure (IPF) colouring and pole figures (Figure~\ref{fig:EBSD}c), the platelets are preferentially oriented with their $c$ axis (001) parallel to Z$_0$ (the pressure direction). 

\begin{figure*}
\centering
\includegraphics[width=17.5cm]{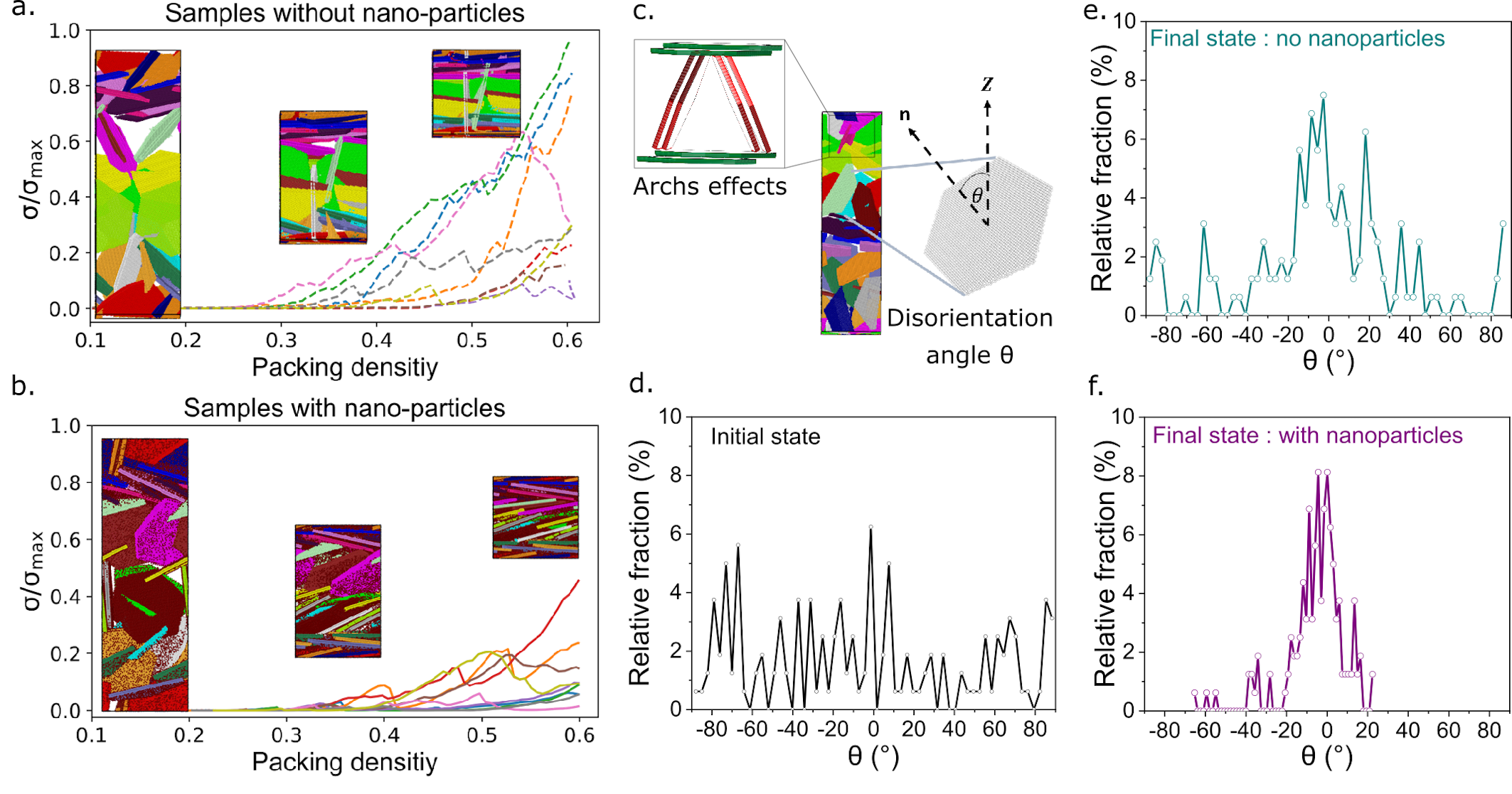}
\caption{\textbf{DEM simulations:} Evolution of the normalized axial stress as a function of the packing density for ten simulations for packings a) without nanoparticles and b) with nanoparticles. The axial stress is normalized by the maximum stress reached for all numerical samples without nanoparticles. c) Initial numerical sample with a schematic representation of the arching effect. Zoom-in on a platelet with its normal vector $\Vec{n}$ and the angle $\theta$ between the pressing direction Z$_0$ and $\Vec{n}$.  d-f) Platelet angle distribution for d) the initial state, the final state e) without nanoparticles and f) with nanoparticles. \textsuperscript{\textcopyright} (2020) H. Saad \emph{et al.} (10.6084/m9.figshare.11907276) CC BY 4.0 license https://creativecommons.org/licenses/by/4.0/.}
\label{fig:DEM_fig}
\end{figure*}

Platelet misalignment was evaluated using the crystalline orientation for each platelet obtained from EBSD~\cite{Brosnian2006}. Figure~\ref{fig:EBSD}d shows the distribution of the disorientation angle $\theta$ (Figure~\ref{fig:DEM_fig}c) in term of relative fraction for the center and the edge (border) of the disc, respectively. The platelets preferential orientation and misalignment are respectively characterized by the mean value of the angular distribution (${\theta}_{mean} =0^{\circ}$ when platelets are perfectly aligned perpendicularly to the pressure direction Z$_0$) and its full-width at half maximum (FWHM). A better alignment was found at the center (${\theta}_{mean} =-0.5^{\circ}$) than at the edges (${\theta}_{mean} =-7^{\circ}$). The misalignment at the center of the sample is evaluated to 30$^{\circ}$ against 40$^{\circ}$ at the edges. This difference is likely due to a friction effect with the die during pressing. The average misalignment is slightly higher compared to what can be obtained using other techniques such as ice-templating (<~20$^{\circ}$), MASC or TGG ($\approx$~10$^{\circ}$) where the platelet alignment is controlled at the microscopic scale~\cite{LeFerrand2019}. The influence of this misalignment on the material mechanical properties, and particularly on toughness, is apparently insignificant which confirms that a perfect alignment is not required to achieve good mechanical properties, and instead that the microscopic local defects (arches, pores) can contribute to the reinforcement by acting as crack propagation barriers. 

\textbf{Platelets alignment during pressing} -- The system self-organizes under the applied pressure during sintering. Several mechanisms might contribute to the alignment. The geometric anisotropy of platelets should favour their shear-induced alignment perpendicular to the applied pressure. The presence of nanoparticles at the surface of platelets could also contribute to their alignment.

We use the Discrete Element Method (DEM)~\cite{Radi2019a,Radi2019b} to investigate numerically the alignment of platelets during pressing (before the melting of the glass phase) and the role of nanoparticles (see \emph{Methods} and \emph{Supplementary informations} for details). Numerical samples containing each twenty randomly oriented platelets with and without nanoparticles were generated. Figures~\ref{fig:DEM_fig}a,b show the evolution of the normalized axial stress in the pressing direction as a function of the packing density of the ten samples used. The normalization is carried out by using the maximum stress value for compacts without nanoparticles. Although stress--density curves are rather scattered, owing to the limited number of platelets, it is clear that for a given packing density, the compacting stress of samples without nanoparticles is larger on average than for those with nanoparticles. The stress peaks followed by stress drops originate from the locking and unlocking of platelets in the packing. 

Platelets may fracture, facilitating rearrangement. This mechanism is not simulated in our DEM simulations. However, experimentally it should be limited since a pressure of only 20~MPa is applied before the melting of the glass phase. Snapshots in figures~\ref{fig:DEM_fig}a,b show the gradual alignment of platelets with the increase of packing density. It indicates that the number of highly misaligned platelets decreases when nanoparticles are present.

The distributions of the angles $\theta$ between the normal vector $\Vec{n}$ of platelets, and the pressing direction (Z$_0$) are plotted in figures~\ref{fig:DEM_fig}d--f. Before pressing, the platelet normal angles are uniformly distributed (Figure~\ref{fig:DEM_fig}d), indicating that platelets are randomly oriented.  For a packing density of 0.6 (typical density prior to the liquid phase formation), the distributions narrow towards $\theta=0^{\circ}$ (well aligned packing). However, this tendency is clearly more pronounced when nanoparticles are introduced as shown by comparing figures~\ref{fig:DEM_fig}e and f. A thorough inspection of platelet rearrangements in the DEM simulations highlights that stable arches are the main cause of the locking of platelets in misaligned orientations (as sketched in Figure~\ref{fig:DEM_fig}c). We observed that the addition of nanoparticles destabilizes these arches by a ball bearing effect, thus favouring alignment. 

The formation of the liquid glass-phase at high temperature further facilitates alignment through sliding between platelets~\cite{German1985}. The molar composition between silica and calcia used here (75:25(SiO$_2$:CaO)) has been chosen because of its high wettability with the surface of the alumina~\cite{Seabaugh1997}. The formation of the secondary phase helps therefore the platelets organization under the applied uniaxial pressure by reducing the friction coefficient that facilitates their shearing but also by generating interfacial capillary forces that contribute to the densification throughout rearrangement mechanisms~\cite{Huppmann1975}. This is also favoured by the rise of the pressure value to 40~MPa during the plateau at 1450$^{\circ}$C to enlarge the contact surface between the aligned platelets and to fill the pores with the molten glass-phase. 

The texture is a consequence of the different mechanisms that take place during pressing and sintering, and will therefore depend on the homogeneity of the spatial distribution of the building blocks. Local disorientation of the platelets can be observed (Figure~\ref{fig:EBSD}a), creating closed porosities. Some arches are also visible on SEM images and we assume that they result from a poor distribution of the glass-phase or an initial packing configuration that does not favour the local alignment of the largest platelets. Local disorders and pores could nevertheless contribute to increase the crack propagation resistance by acting as local crack arresters.  

\textbf {Sample size effects on the $R-$curves} -- The existence of a $R-$curve in ceramic materials arise from the presence of extrinsic toughening mechanisms, such as crack bridging by large grains. This behaviour in ceramics is known to be a geometry-dependant property with influence of the loading configuration~\cite{Steinbrech1990} and of the sample size~\cite{Lutz1995}. In nacre-like ceramics, the shape of the $R-$curve can greatly vary from one material to another with either an abrupt rise followed by a plateau~\cite{Munch2008, Launey2009a, Launey2010} or a quasi-linear toughness increase with crack length~\cite{Bouville2014, LeFerrand2015}. Effects associated to the specimen dimensions arise during crack propagation when the uncracked ligament length becomes too small compared to the size of the fracture process zone (FPZ). Such effects must therefore be taken into account, especially when they affect the shape of the $R-$curve and can skew the measurement of the maximum fracture toughness.

The ASTM-E1820 standard, initially defined for materials exhibiting plasticity, takes into account these effects by limiting the validity domain of toughness calculation using SENB to a crack extension value, proportional to the sample thickness (${\Delta}a_{max}=0.25 \times b_0$). In order to evaluate sample size effects on the $R-$curves, SENB tests were performed on specimens with thickness varying from 2~mm to 10~mm (Figure~\ref{fig:SENB_bulk}a). For all geometries, we kept a constant initial ratio $\frac{a_0}{W}=0.5$ (where $a_0$ is the initial notch length and $W$ the sample thickness, see \textit{Methods}). Increasing the thickness from 2~mm to 10~mm delays the apparition of these size effects and shifts the ASTM crack extension limit to higher values, corresponding to an increase of the validity domain of the $R-$curve from less than 0.3~mm to more than 1.2~mm. The maximum valid fracture toughness (ASTM standard limit) increases with the sample thickness, from 14.4 to 17.6~MPa.m$^{1/2}$ (see Table~S1).

\begin{figure}
\centering
\includegraphics[width=8cm]{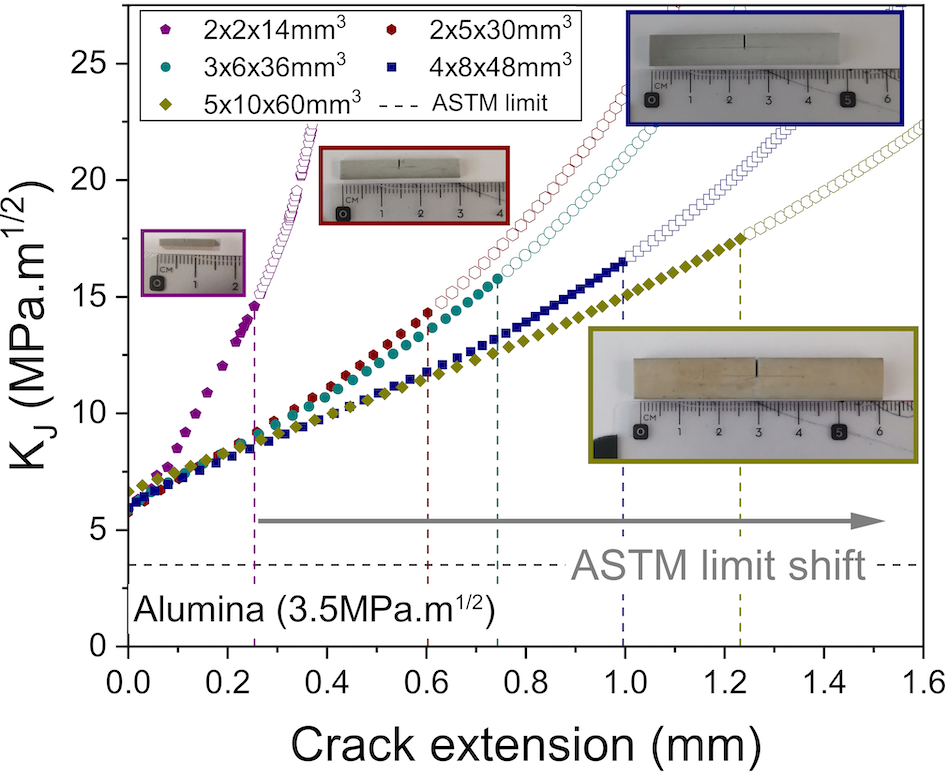}
\caption{$R-$curves for increasing sample thickness: hollow markers correspond to points beyond the ASTM limit, measurements are not valid above this limit. The limit is shifted to the right when the sample thickness increases. \textsuperscript{\textcopyright} (2020) H. Saad \emph{et al.} (10.6084/m9.figshare.11907276) CC BY 4.0 license https://creativecommons.org/licenses/by/4.0/.}
\label{fig:R-curve-2}
\end{figure}

As expected, all the geometries exhibit $R-$curves with a similar crack initiation toughness $K_{J0}$ comprised between 5.5 and 6.6~MPa.m$^{1/2}$ followed by a quasi-linear increase of $K_J$ as the crack extends. For small crack extension values, all the curves follow an envelope curve associated to the 1~cm thick sample. The resistance to crack initiation ($K_{J0}$) is probably the most critical parameter for the design of ceramic components, defining a threshold value below which no crack propagation occurs. According to figure~\ref{fig:R-curve-2}, $K_{J0}$ is an intrinsic material property that does not depend on sample dimensions (see Table~S1). These results are also in good accordance with previous studies performed on 2mm thick nacre-like alumina samples elaborated by freeze-casting (6.2~MPa.m$^{1/2}$~\cite{Bouville2014}) or by magnetic-assisted slip casting (6~MPa.m$^{1/2}$~\cite{LeFerrand2015}) and confirm the robustness of these peculiar microstructures. 

Figure~\ref{fig:R-curve-2} also shows the apparition of size effects during crack propagation for the various geometries, with a deviation from the envelope curve (the $R-$curve of the 1~cm thick sample) that occurs later as the sample thickness increases from 2~mm to 8~mm. Such effects associated with the size of the fracture process zone have been documented for a long time~\cite{Lutz1995}. The sample size effect on the $R-$curve has also been studied in cement based materials for which it is easier to process large samples (from centimeters to a hundred of centimeters). \textit{Cotterell et al.} showed, using SENB tests, that the development of the fracture process zone during the crack extension had a more significant influence on the compliance change of small cement paste specimens leading therefore to a steeper rise of the $R-$curve as compared to larger samples~\cite{Cotterell1987}. The mechanisms associated to the macroscopic material toughening in cement based materials and nacre-like materials are nevertheless different. The damage behavior of concrete results mainly of microcracking mechanisms acting in front of the crack tip in a large fracture process zone. Our nacre-like system are characterized by an important crack deviation (80$^{\circ}$) from the notch plane at initiation, which results from the presence of interfaces perpendicularly to the crack direction. 

\textbf{Perspectives} -- The simplified process described here, based on uniaxial pressing, may enable the scale-up and industrial production of nacre-like damage-tolerant ceramic materials. We demonstrated here that a sufficient pressure can allow the long-range alignment of a system containing anisotropic platelets that can lead to nacre-like microstructure. Although the quality of platelets alignment is not as good as with alternative processes reported previously, this does not seem to impact the mechanical properties. Such bioinspired brick-and-mortar microstructures are therefore apparently very robust. The availability of industrial tools should facilitate the investigation and development of these bioinspired ceramic/ceramic composites for structural applications. Future work will focus on the control of the interface, which is critical to optimize the mechanical properties.

\section*{Experimental section}

\emph{Suspensions preparation} -- The suspensions were prepared following a method used previously~\cite{Bouville2014}. The system is composed of micron-sized alumina platelets, alumina nanoparticles and glass-phase precursors : calcium carbonate and colloidal silica. The suspensions were then ball-milled, quenched in liquid nitrogen and then freeze-dried. The obtained powder was thermally treated to remove the organic compounds.

\emph{Field-assisted Sintering (FAST)} -- Field-assisted sintering (FAST) was performed using graphite dies, under vacuum. The selected heating cycle was 25$^{\circ}$C/min up to 1450$^{\circ}$C under a constant uniaxial pressure of 20~MPa followed by a plateau of 15min at 1450$^{\circ}$C and under a pressure of 40~MPa. 

\emph{DEM simulations} -- The in-house code dp3D was used to generate initial random packings of platelets and to press them unidirectionally.  Ten initial configurations were tested starting from different initial random seeds. The exact same configuration of platelets was used for packings with or without nanoparticles. A non-smooth surface was obtained when meshing the platelets with spheres, providing some resistance to sliding. Thus, it was found unnecessary to introduce friction between contacts. 

\emph{Microstructural characterization} -- The microstructure of the material was evaluated using scanning electron microscopy on uncoated mirror polished samples. Electron backscattered diffraction (EBSD) was used to generate crystallographic orientation maps.

\emph{Mechanical characterization} -- Rectangular beams were cut from the sintered disc for the mechanical characterization of the material. For each set-up, at least 5 samples were tested. The flexural strength was measured by both three-points and four-points bending tests, according to the ASTM E1820-E01 standard. Single-edge notched beam (SENB) tests were performed to characterize the fracture toughness of the material and its crack-resistance curve ($R-$curve). The compliance method ($C=\frac{d}{F}$) was used here to determine the beginning of the crack propagation and to determine the projected crack length. Non-linear elastic fracture mechanisms analysis was used to determine the crack-resistance curves of the material ($R$-curves). $R$-curves were measured in terms of the $J-$integral as a function of the crack extension. Toughness parameters such as crack initiation toughness (K$_{I0}$) and fracture toughness ($K_{IC}$) were directly determined from the $R-$curves. The Young's modulus $E$ and the Poisson's ratio $\nu$ of the material were determined experimentally using ultrasonic contact technique.

\section*{Supporting Informations}
Detailed methods, table S1.

\section*{Acknowledgements}
We acknowledge the technical support of Saint-Gobain Research Provence. The Agence Nationale pour la Recherche (ANR-16-CE08-0006 BICUIT, program DS0303-2016) and the AGIR-POLE-PEM 2016 project (OMicroN) are greatly acknowledged for their financial support.

\section*{Conflict of interest}
The authors declare no conflict of interest.

\bibliographystyle{apsrev4-1} 
\bibliography{reference.bib}

\end{document}